\documentclass[reprint,groupedaddress,showpacs]{revtex4-1}

\usepackage{dcolumn}
\usepackage{amsfonts,amsmath,amssymb}
\usepackage{appendix}
\usepackage[latin1]{inputenc}
\usepackage{color}
\usepackage{ulem}
\usepackage{graphicx}

\graphicspath{fig/}

\definecolor{forestgreen}{RGB}{34, 139, 34}

\begin{document}

\title{Large 2D Coulomb crystals in a radio frequency surface ion trap}

\author{B. Szymanski}
\email{benjamin.szymanski@univ-paris-diderot.fr}
\author{R. Dubessy}
\altaffiliation[Present address ]{Laboratoire de Physique des Lasers, CNRS-UMR7538, Universit{\'e} Paris 13--Institut Galil{\'e}e, Villetaneuse, France }
\author{B. Dubost}
\altaffiliation[Also at ]{ICFO-Institut de Ciencies Fotoniques, Mediterranean Technology Park, 08860 Castelldefels (Barcelona), Spain}
\author{S. Guibal}
\author{J.-P. Likforman} 
\author{L. Guidoni}
\affiliation{Univ Paris Diderot, Sorbonne Paris Cit{\'e}, Laboratoire Mat{\'e}riaux et Ph{\'e}nom{\`e}nes Quantiques, UMR 7162, Bat. Condorcet, 75205 Paris Cedex 13, France}

\date{\today}


\begin{abstract} 
We designed and operated a surface ion trap, with an ion-substrate distance of 500~$\mu$m, realized with standard printed-circuit-board techniques. The trap has been loaded with up to a few thousand Sr$^+$ ions in the Coulomb-crystal regime.
An analytical model of the pseudo-potential allowed us to determine the parameters that drive the trap into anisotropic regimes in which we obtain large ($N>150$) purely 2D ion Coulomb crystals.
These crystals may open a simple and reliable way to experiments on quantum simulations of large 2D systems.
\end{abstract}

\maketitle

Applications of laser cooled trapped ions are numerous and cover several research fields such as quantum information processing \cite{Cirac:1995,Leibfried:2003,Haffner:2008}, quantum simulation \cite{Porras:2004a,Johanning:2009}, cold molecule spectroscopy \cite{Willitsch:2010} and metrology \cite{Margolis:2010}.
In the frame of quantum information processing, a large-scale computer architecture has been proposed based on ion shuttling between interaction and memory zones \cite{Kielpinski:2002}.
A very practical way to realize such an architecture relies on surface electrode radio-frequency (rf) traps \cite{Chiaverini:2005,Seidelin:2006} in which a pseudo-potential well is created above the surface of a substrate by a set of deposited metallic electrodes.
The vast majority of the surface traps developed so far \cite{Britton:2009,Moehring:2011} are devoted to the trapping and shuttling of short ion strings (1D ion Coulomb crystals). 
However a peculiar characteristic of the planar geometry, never exploited to date, is its intrinsic anisotropy that can lead to the creation of single-layer 2D ion Coulomb crystals.
As suggested by Porras and Cirac \cite{Porras:2004}, such crystals are well adapted to simulate quantum phase transitions in spin systems. 
In particular, the control of anisotropy is ideally suited for the study of zigzag transition instabilities directly related to one-dimensional Ising models in a transverse field \cite{Shimshoni:2011,Bermudez:2011}.

In order to create purely 2D laser-cooled ion lattices, three different strategies have been considered so far.
One relies on Penning traps. It allowed for the observation of structural phase transitions \cite{Mitchell:1998} and has been more recently used for quantum control experiments \cite{Biercuk:2009}.
The main disadvantage of this strategy is the difficulty of laser-cooling that has to deal with both magnetron and cyclotron motions \cite{Biercuk:2009}.
\\
\noindent Radio-frequency trap arrays are also a very promising scheme \cite{Schmied:2009}. 
The main advantage of this strategy is the possibility to design regular lattice structures that are not imposed by the self-arrangement.
However, the experimental inter-ion distances are, up to now, quite large and do not allow for sufficient ion-ion interaction  \cite{Clark:2009}.
\\
\noindent Finally, the surface point Paul trap geometry \cite{Pearson:2006a}, ideally suited to vary the trapping distance from the substrate, can be a promising candidate for quantum simulation \cite{Kim:2010}. Up to ten ions in a 2D cristalline arrangement have been loaded in this kind of traps realized with printed circuit board technology and operated in a cryogenic environment\cite{Clark:2011}.

In this paper we present a linear surface rf ion trap based on a standard printed circuit board and we demonstrate the versatility of such a device that allows for the trapping of large crystallized ion ensembles.
Depending on the trap parameters, different crystal shapes can be obtained.
In particular, we demonstrated the formation of single-layer Coulomb crystals containing more than 150 ions.
Previous works have proposed and realized similar devices \cite{Brown:2007, Kim:2010} in which the presence of stray fields \cite{Brown:2007} or the chosen trap geometry \cite{Kim:2010} probably prevented the formation of large Coulomb crystals. 
\\

\begin{figure}[h]

\includegraphics[width=.7\columnwidth]{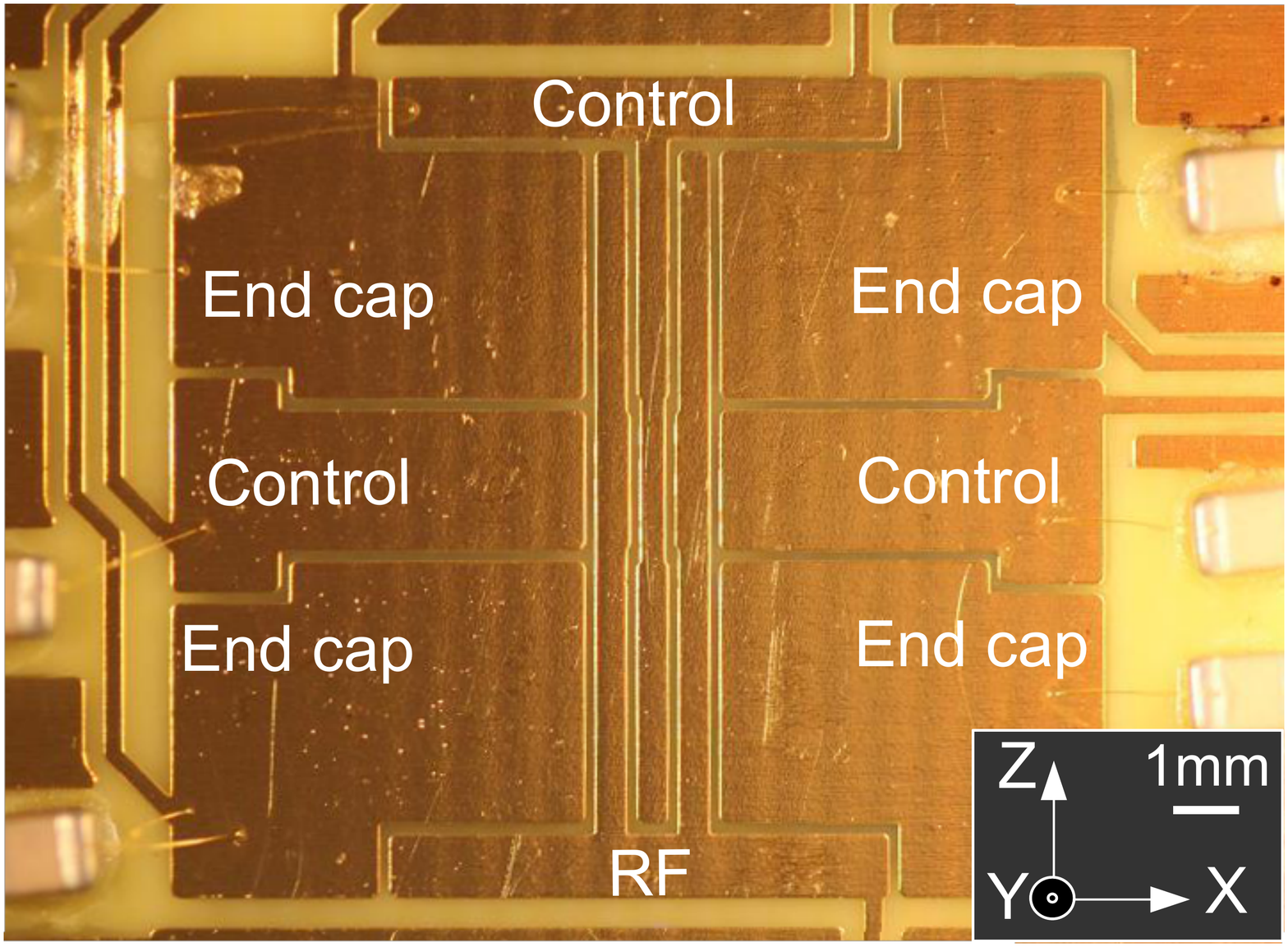}
  \caption{Picture of the gold-plated printed circuit board surface trap realized by standard commercial technique on a FR4 card without any further operation. The dimensions are: 500~$\mu$m wide rf tracks and 340~$\mu$m wide central electrode. The inter-electrode spacing is 200~$\mu$m. To minimize the rf coupling, surface-mount capacitors (10 nF) (visible on the sides of the picture) are bonded between each dc electrode and the ground plane.}
  \label{fig:scheme_trap}
\end{figure}

We used a copper FR4 printed circuit board on which strip-lines, forming the ion trap electrodes, were chemically etched and gold-plated (thickness $<1$~$\mu$m) using standard commercial procedures.
The board material has already proven to be UHV compatible and bakable up to $150^{\circ}$C~\cite{Rouki:2003}.
The five wire trap geometry is presented in Fig.~\ref{fig:scheme_trap}. The longitudinal confinement is assured by four "end cap'' electrodes. An oscillating potential $V_{rf}\cos(\omega t)$ with typically $V_{rf}=125$~V and $\omega/2\pi= 6.9$~MHz is applied to the rf electrodes.
Static voltages in the range -5V to +5V are typically used to drive the central control electrode ($V_{CC}$), the two lateral control electrodes ($V_{LC}$) and the four endcaps ($V_{EC}$).

\begin{figure}[h]
\centerline{\hfil \includegraphics[width=1\columnwidth]{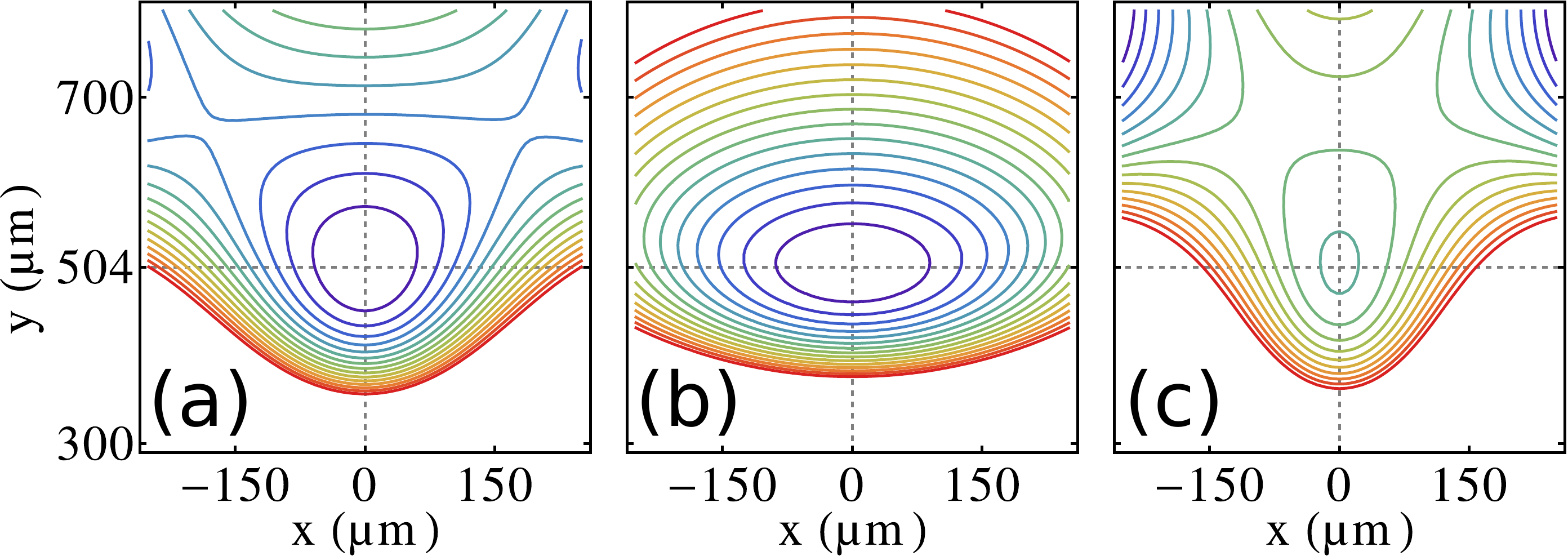}}
\caption{Pseudo-potential cross section calculated for three sets of trap parameters. In all cases the ion-surface distance is 504~$\mu$m, fixed by the trap geometry and $q{_x}$=0.173, $q{_y}$=-0.171 and $q{_z}$=-0.002. 
(a) isotropic potential obtained for  $V_{rf}$=125~V, $V_{CC}$=-2.25~V, $V_{LC}$=-11~V and $V_{EC}$=5~V. The trap depth is 38~meV 
and the motional frequencies are $\omega{_x}$=408~kHz, $\omega{_y}$=404~kHz,  $\omega{_z}$=156~kHz. The stability parameters are $a_x$=-0.001, $a_y$=-0.001 and $a_z$=-0.002.
(b) anisotropic potential obtained for  $V_{rf}$=125~V, $V_{CC}$=3.73~V, $V_{LC}$=5~V and $V_{EC}$=5~V. The trap depth is 141~meV and the motional frequencies are $\omega{_x}$=266kHz,  $\omega{_y}$=529kHz,  $\omega{_z}$=39kHz. The stability parameters are $a_x$=-0.009, $a_y$=-0.009 and $a_z$=0.0001.
(c) anisotropic potential obtained for  $V_{rf}$=125~V, $V_{CC}$=-6~V, $V_{LC}$=5~V and $V_{EC}$=-21.05~V. The trap depth is 13~meV and the motional frequencies are $\omega{_x}$=474kHz,  $\omega{_y}$=297kHz,  $\omega{_z}$=197kHz. The stability parameters are $a_x$=0.0039, $a_y$=-0.0072 and $a_z$=0.0033.
Isopotential curves are separated by 10~meV.}
\label{fig:cal_potentials}
\end{figure}

Following reference \onlinecite{House:2008}, we performed an analytical calculation of the pseudo-potential associated to this particular trap geometry.
The calculation gives us useful information such as the ion motional frequencies as a function of the trap parameters $V_{rf}$, $V_{CC}$, $V_{LC}$ and $V_{EC}$, thus determining the trap axial and transverse anisotropies.
The ion distance from the trap surface (504$\mu$m, imposed by the geometry), the trap depth and the stability parameters are also obtained. Using an approach similar to that described in reference  \cite{Ozakin:2011}, we can also calculate the generalized $q$ and $a$ stability parameters: $q_x$, $q_y$, $q_z$ $a_x$, $a_y$ and $a_z$. As an example, Fig.~\ref{fig:cal_potentials} shows three theoretical radial ($xy$) cross sections of the pseudo-potential obtained with three sets of trapping parameters that produce anisotropic or isotropic potentials.

Sr$^+$ ions are created in the trapping region (typical rate $\sim 20$~s$^{-1}$) out of an atomic vapor using a photoionization technique based on two-photon absorption of femtosecond pulses \cite{Removille:2009}.
The ions are Doppler cooled using the 711~THz  $5 ^2S_{1/2}\to 5 ^2P_{1/2}$ optical transition ($\lambda=422$~nm).
To avoid optical pumping into the metastable $4^2D_{3/2}$ state we use an additional laser adressing the 275~THz $4^2D_{3/2}\to 5 ^2P_{1/2}$  transition ($\lambda=1092$~nm).
The laser set-up is very similar to the one described in reference \onlinecite{Removille:2009a}.
The trap is placed in a UHV chamber with an estimated pressure below $10^{-9}$~mbar.

In the experiment, voltages slightly different from the ideal symmetric case have to be applied to the control electrodes in order to position a single trapped ion precisely at the node of the rf electric field.
This allows for the reduction of the micro-motion of the ion driven by the rf electric field.
In order to optimize these voltages, we used the rf correlation technique \cite{Berkeland:1998} that measures the arrival time correlations of  single fluorescence photons with the rf cycle.
By using purely longitudinal ($z$-propagating) and transverse (propagating at $45^{\circ}$ in the $xz$ plane) cooling beams we were able to compensate for a residual axial micro-motion and the micro-motion along $x$.
In order to compensate for the micro-motion along $y$ (vertical), we implemented the technique developed by Allcock and co-workers, based on a vertical repumping beam \cite{Allcock:2010}.

We observe stable trapping and micro-motion compensation voltages which do not vary significantly over a 30~min time-scale.
However, optimal compensation voltages are not stable on a day-to-day basis (typical drifts of 70 V/m).
With optimized compensation voltages, the single ion lifetime is $\sim 20$~min, probably limited by the pressure in the vacuum chamber.
Single ion fluorescence spectra confirm a linewidth limited by the lifetime of the $5 ^2P_{1/2}$ state ($\Gamma/2\pi=21.7$~MHz).

\begin{figure}[h]
 \centerline{\includegraphics[width=.75\columnwidth]{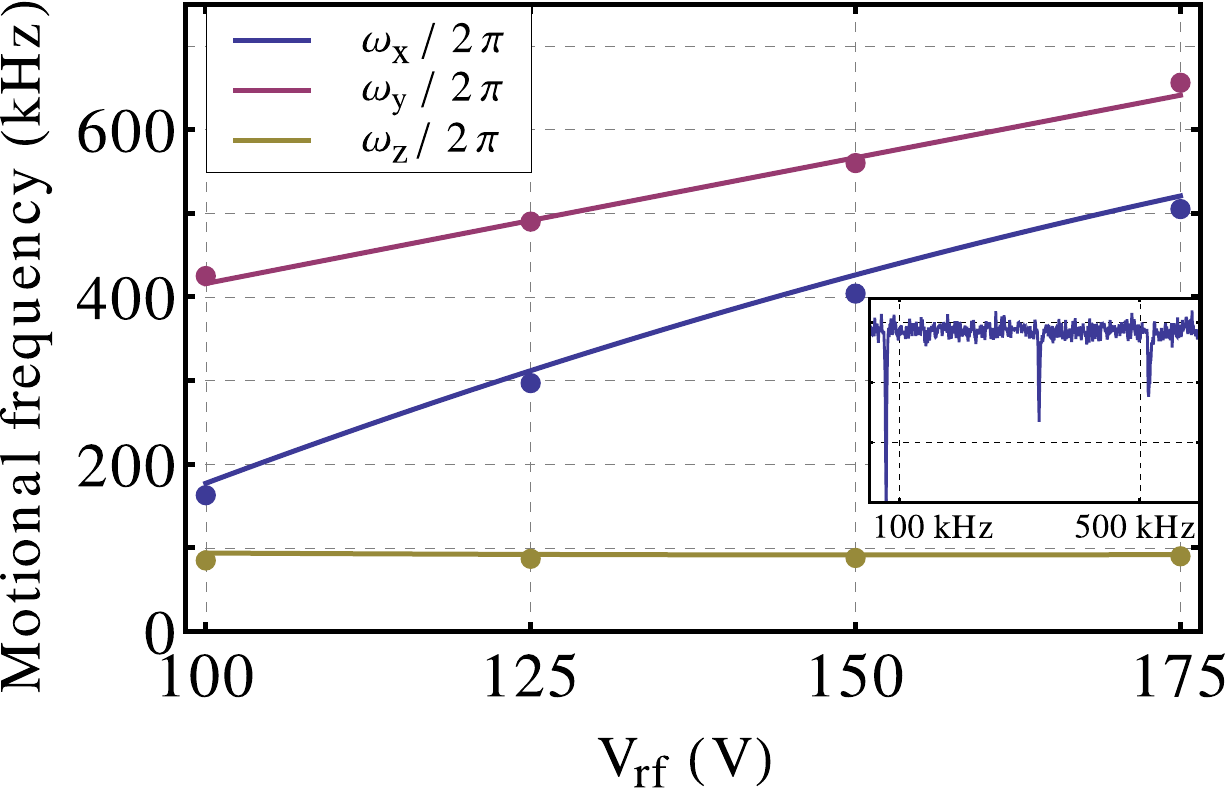}}
  \caption{Measured (dots) and calculated (lines) motional frequencies as function of $V_{rf}$. The inset shows an example of an excitation spectrum i.e. single-ion fluorescence intensity as a function of excitation frequency at a fixed $V_{rf}$.}
  \label{fig:fig3_tickle}
\end{figure}

A first test of the pseudo-potential calculation is performed by comparing the experimentally measured motional frequencies with an harmonic fit of the calculated pseudo-potential at the trap center. The motional frequencies are obtained by measuring the single-ion fluorescence intensity as function of the frequency of the excitation voltage (tickle) applied to an endcap electrode (see inset of Fig.~\ref{fig:fig3_tickle}). The results, shown in  Fig.~\ref{fig:fig3_tickle}, present a very good agreement between calculation and experiment.

\begin{figure}[h]
 \centerline{\includegraphics[width=1\columnwidth]{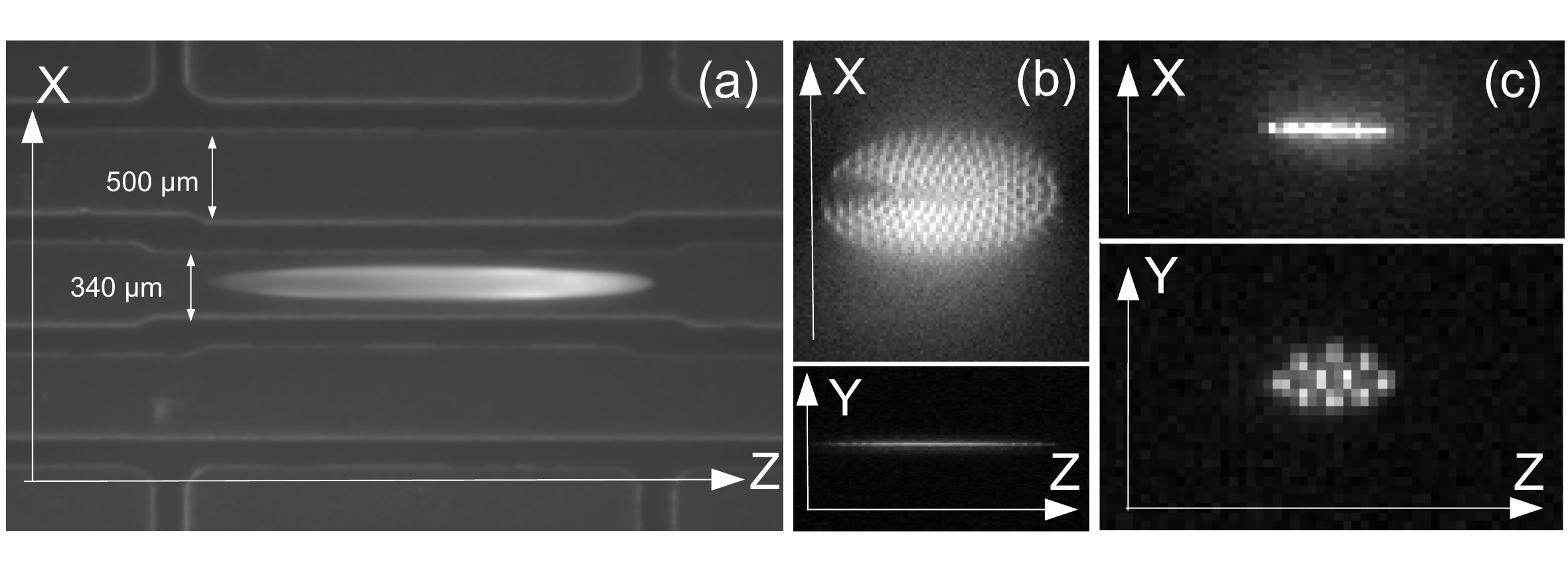}}
  \caption{Fluorescence images of the trapped ions. (a) top-view of a large 3D Coulomb crystal containing $\sim 4500$~ions. 
  (b) 2D Coulomb crystal containing $\sim 150$~ions arranged in a plane parallel to the trap surface (the inter-ion distance is 11~$\mu$m). The single-layer character is evidenced by the lateral view.
  (c) Single-layer crystal arranged in the $yz$ plane perpendicular to the printed circuit board (the inter-ion distance is 9~$\mu$m).}
  \label{fig:crystals}
\end{figure}

Typical fluorescence images of the trapped ions obtained for different trap parameters are shown in  Figure~\ref{fig:crystals}. In Fig.~\ref{fig:crystals}(a) the ions organize themselves as a large three dimensional Coulomb crystal. We observed 3D crystals containing up to a few thousand ions, comparable to the typical numbers obtained in three dimensional macroscopic linear Paul traps \cite{Drewsen:1998}.

Another and particularly interesting configuration is the single layer Coulomb crystal, demonstrated in Fig.~\ref{fig:crystals}(b).
Using an imaging system aligned along the $x$ direction, we have actually checked the single layer character of this Coulomb crystal (bottom image).
The non-fluorescing ions visible on the top view of the crystal (left side) are sympathetically-cooled strontium isotopes not addressed by the cooling lasers (only $^{88}$Sr$^+$ is laser-cooled in this experiment).
Trap potential calculations allowed us to find a very unusual working point in which the Coulomb crystal forms a single layer perpendicular to the trap surface, as shown in Fig.~\ref{fig:crystals}(c). However, contrary to the two previous cases, ions could not be directly loaded using the calculated parameters (see caption of Fig.~2c for this case), propably due to the small value of the trap depth (13~meV). 
The trap is actually loaded using parameters close to those used in Fig.~\ref{fig:cal_potentials}(c) but with a higher value of $V_{CC}$, in order to increase the trap depth. Then, $V_{CC}$ is reduced and one can observe the formation of a mono-layer ion crystal perpendicular to the trap surface.With the current trap design we could actually trap up to sixteen ions in this configuration. 
As mentioned above, these 2D structures may be exploited for the quantum simulation of two dimensional systems \cite{Porras:2006}. In particular, the "vertical'' arrangement could allow for an easier ion addressing by lasers since the control beams could freely propagate parallel to the trap surface.
\\


In this work, we have demonstrated the versatility offered by an inexpensive, easily fabricated rf surface ion trap based on a printed-circuit board.
We have shown that large Sr$^+$ ion crystals arranged in a 3D structure can be obtained, comparable in size to the crystals used in recent cavity quantum electrodynamics experiments\cite{Herskind:2009a}.
The same device allowed us to create large ion Coulomb crystals purely bi-dimensional lying parallel to the trap substrate and containing up to 150 ions.
In addition, we also demonstrated 2D Coulomb crystals standing in a plane perpendicular to the trap surface.
This particular geometry allows for an easier individual addressing of ions, especially useful in view of quantum simulation experiments.
This kind of versatile devices will probably become a practical tool for quantum simulation experiments.
\\

We thank M. Apfel and P. Lepert for technical support.
We would like to thank D. T. C. Allcock for fruitful discussions and suggestions.
We acknowledge financial support by R{\'e}gion Ile-de-France through the SESAME project. B. S. gratefully acknowledges the funding from the D{\'e}l{\'e}gation G{\'e}n{\'e}rale de l'Armement and the French Ministry of Education and Research.

\bibliography{biblio_total}

\end{document}